\documentclass[aps,pre,onecolumn,superscriptaddress,nofootinbib]{revtex4}

\usepackage{amsmath,amsfonts,amssymb,amsthm}
\usepackage{overpic}
\usepackage{mhchem}
\usepackage{bm}
\usepackage{graphicx}
\usepackage[colorlinks=true,linkcolor=blue,citecolor=blue,urlcolor=blue]{hyperref}

\newcommand\dd{\mathrm{d}}
\newcommand\n{\mathbf{n}}

\newcommand\vvec{\bm{v}}

\newcommand\x{\bm{x}}

\newcommand\X{\mathbf{X}}
\newcommand\Qvec{\bm{Q}}

\newcommand\pp{\partial}

\newcommand\uvec{\bm{u}}

\newtheorem{rmk}{Remark}[section]

\begin{document}

\title{Energetic Variational Modeling of Active Nematics: Coupling the Toner--Tu Model with ATP Hydrolysis}

\author{Yiwei Wang}
\affiliation{Department of Mathematics, University of California, Riverside, Riverside, CA 92507, USA}
\email{yiweiw@ucr.edu}

\begin{abstract}
We present a thermodynamically consistent energetic variational model for active nematics driven by ATP hydrolysis, with a focus on the coupling between chemical reactions and mechanical dynamics. Extending the classical Toner--Tu framework, we introduce a chemo-mechanical coupling mechanism in which the self-advection and polarization dynamics are modulated by the ATP hydrolysis rate. The model is derived using an energetic variational approach that integrates both chemical free energy and mechanical energy into a unified energy-dissipation law. The reaction rate equation explicitly incorporates mechanical feedback, revealing how active transport and alignment interactions influence chemical fluxes and vice versa. This formulation not only preserves consistency with nonequilibrium thermodynamics but also provides a transparent pathway for modeling energy transduction in active systems. We also present numerical simulations demonstrating how ATP consumption drives the merging of topological defects and enables the system to escape a quasi-equilibrium—a phenomenon not observed in passive nematic systems. This framework offers new insights into energy transduction and regulation mechanisms in biologically related active systems.
\end{abstract}

\maketitle


\section{Introduction}

Active matter refers to a class of materials composed of active units that convert chemical energy into mechanical work \cite{das2020introduction, doostmohammadi2022physics, marchetti2013hydrodynamics, needleman2017active}. Examples of active matter include
living and non-living systems, such as collections of molecular motor, cells, bacteria, synthetic microswimmers, and self-propelled colloids \cite{aranson2022bacterial, saintillan2008instabilities, sanchez2012spontaneous, wu2011self}. In recent years, active matter has gained significant attention due to their potential for various biomedical applications, such as drug delivery and soft robotics \cite{cianchetti2018biomedical, ozkale2021active, shah2018stimuli}.

One of the fundamental goal in studying of active matter is to obtain a quantitative, mechanistic understanding of various chemo-mechanical processes in different spatial-temporal scales in these systems \cite{das2020introduction}. For example, 1) how mechanical properties like viscoelasticity and stiffness are affected by biological, chemical, and physical cues such as cell-cell contacts, biochemical signaling, temperature, and voltage \cite{alber2019integrating, liu2022living}. 
2) how to quantify the energy conversion, entropy production and heat generation \cite{das2020introduction, o2022time}. 
Building computational models for active matter is an indispensable way to achieve this goal.
Despite many mathematical models of active materials have been developed \cite{marchetti2013hydrodynamics, shaebani2020computational}, most existing models are developed by adding an active stress or force to existing passive material models. The detailed biochemical interaction, such as ATP hydrolysis, are not included explicitly in these model.  
As a consequence, these models may have limited insights into the origins of activity and responsively of active materials.

Based on fundamental principles of non-equilibrium thermodynamics \cite{de2013non, kjelstrup2008non}, Energetic Variational Approach (EnVarA) \cite{liu2009introduction, giga2017variational, wang2022some}, together with other variational principles such as Onsager's variational principle \cite{doi2011onsager, wang2021generalized}, General Equation for the Nonequilibrium Reversible-Irreversible Coupling (GENERIC) \cite{grmela1997dynamics, ottinger1997dynamics}, offer a promising way to develop better models for active matter These variational principles have been successfully applied in developing thermodynamically consistent models for passive soft matter including chemo-mechanical coupling \cite{wang2020field} and non-isothermal effects \cite{liu2021brinkman}. 
The main idea of these approaches is to derive a mathematical model of a complex system based on variations of a free energy and a dissipation functional that related to the rate of entropy production \cite{wang2022some}. 
Through the choice of the free energy and the dissipation,  the coupling and completion of various multiscale, multiphysics, chemo-mechanical processes can be included, enabling the quantification of energy consumption, the rate of entropy production, and heat generation in the developed model.

However, extending these variational principles to active matter is challenging as they were originally developed for closed and dissipative systems. The functionality of active matter requires continuously injection of energy through chemical energy, mechanical force, electric fields, or light \cite{das2020introduction, morrow2019chemically, o2022time}
to maintain the out--of--equilibrium states, they need to be viewed as either open systems or a subsystem for a significant large close system \cite{ge2013dissipation}. Although substantial efforts have been made in this direction recently \cite{ackermann2023onsager,klamser2018thermodynamic, mirza2025variational, markovich2021thermodynamics, nardini2017entropy, o2022time, wang2021onsager}, further development of modeling and computational tools is still needed to gain deeper insights on active materials. 

The purpose of this paper is to explore the possibility of constructing a thermodynamically consistent model for active matter using the energetic variational approach. The key idea is to explicitly incorporate the chemical reactions that power the system, i.e., ATP hydrolysis, into the model. Although the resulting chemo-mechanical system is dissipative, i.e., the total free energy decreases over time, the mechanical free energy may increase due to energy transduction from chemical to mechanical forms. Existing active matter models can be recovered as special cases by prescribing a constant ATP hydrolysis rate, which represents continuous energy input. To illustrate this new modeling framework, we focus on a variant of the Toner--Tu model \cite{toner1995long}, which describes the large-scale dynamics of self-propelled particles via coarse-grained velocity and density fields. The developed approach can be used to revise more complex active soft matter systems. This variational framework provides a new perspective for modeling active matter in a thermodynamically consistent manner and enables a quantitative description of energy transduction.

The remainder of the paper is organized as follows. In Section 2, we briefly review the classical Toner--Tu model and its variants, drawing connections to nematic liquid crystals. We also introduce the framework of EnVarA and discuss why Toner--Tu-type models cannot be directly derived from this framework. In Section 3, we develop a new thermodynamically consistent chemo-mechanical model. Numerical simulation results are presented in Section 4. Finally, concluding remarks are provided in Section 5.

\section{Preliminaries}

\subsection{The Toner--Tu model and its variants}

In this section, we briefly review the Toner--Tu (TT) model and its variants. The TT model can be viewed as a continuum-level description inspired by the Vicsek model \cite{vicsek1995novel}, which captures the collective behavior of self-propelled particles with alignment interactions. 
In the Vicsek model, each particle $i$ at position $\x_i(t)$ moves with a constant speed $v_0$ in the direction of its heading ${\bm \theta}_i(t)$. The direction of moving is updated at each time step according to 
 \begin{equation}
 {\bm \theta}_i (t+1) = \langle {\bm \theta}_i(t) \rangle_R  + {\bm \xi}_i (t)
 \end{equation}
 where, $ \langle {\bm \theta}_i(t) \rangle_R$ is the average orientation vector of particles located in a circle of radius $R$ surrounding particle, ${\bm \xi}_i$ is the a random vector with orientation obtained from a uniform distribution $[-\sigma \pi, \sigma \pi]$ with $\sigma$ being the strength of the noise. The new poistion of particle $i$ is updated by
 \begin{equation} 
 \x_i(t + 1) = \x_i (t) + v_0 \hat{\bm \theta}_i
\end{equation}
where $\hat{\bm \theta}_i$ is the unit vector in the direction of ${\bm \theta}_i$. Mathematically, the Vicsek model can be written as \cite{bolley2012mean}
 \begin{equation}
  \begin{cases}
    & \dd X_t^i = \dd V_t^i \dd  t \\
    & \dd V_t^i =  - P(V_t^i) \left( \frac{1}{N} K(X_t^i - X_t^j) (V_t^i - V_t^j)\right) \dd t + \sqrt{2} P(V_t^i) \circ \dd W_t
  \end{cases}
 \end{equation}
 where $P(\vvec)$ is the projection operator on the tangent space at $\frac{\vvec}{|\vvec|}$ to the unit sphere, given by
 \begin{equation}
  P(\vvec) = I  - \frac{\vvec \otimes \vvec}{|\vvec|^2}
 \end{equation}
 and $K$ is the nteraction kernel. In the orginal Vicsek model,
 \begin{equation}
 K(x) =  \begin{cases}
  & 1, \quad |x| \leq R \\
  & 0, \quad |x| > R. \\
 \end{cases}
 \end{equation}
 The term $\sqrt{2} P(V_t^i) \circ \dd W_t$ introduces angular noise into the dynamics. Here, $\dd W_t$ is a standard Brownian motion in velocity space, and the Stratonovich integral (denoted by $\circ \dd W_t$) preserves the geometric structure.
 The activity in the Vicsek model is encoded in the self-propulsion at constant speed $v_0$ and the nonequilibrium dynamics of heading update. This continuous energy input leads to rich collective behavior, such as flocking and phase separation, not seen in equilibrium systems \cite{vicsek1995novel, ginelli2016physics}.
 
Although the Toner--Tu model is not derived as a formal hydrodynamic limit of the Vicsek model, it is built by leveraging its key symmetries and conservation laws \cite{toner1995long}. The original Toner--Tu model, proposed in \cite{toner1995long}, is formulated as 
\begin{equation}\label{TT_orig}
  \begin{cases}
    & \pp_t n + \nabla \cdot (n {\bm v}) = 0 \\
    & \pp_t {\bm v} +  \lambda_1 ({\bm v} \cdot \nabla ) {\bm v}  = (\alpha  - \beta |{\bm v}|^2) {\bm v} - \nabla P + D_L \nabla (\nabla \cdot {\bm v}) + D_1 \Delta {\bm v} + D_2 ({\bm v} \cdot \nabla )^2 {\bm v} + {\bm f}
  \end{cases}
\end{equation}
where $n(\x, t)$ is the number density of active particles, and ${\bm v}$ is the velocity or the polarization field, ${\bm f}$ is the Gaussian random noise. The phenomenological parameters $\beta, D_1, D_2$ and $D_L$ are all positive, while $\alpha < 0$ is the discordered phase and $\alpha > 0$ in the ordered state. In the ordered phase, the $\alpha$ and $\beta$ terms simply make the local ${\bm v}$ have a nonzero magnitude $\sqrt{\alpha / \beta}$  in the ordered phase, corresponding to the constant speed assumption in the orginal Vicsek model.
The pressure $P$ is assumed to be a function of density, given by
\begin{equation}
  P(n) = \sum_{\alpha=1}^{\infty} \sigma_{\alpha} (n - n_0)^{\alpha}
\end{equation}
where $n_0$ is the mean of the local number density and $\sigma_{\alpha}$ are coefficients in the pressure expansion.

In the special case where \(\lambda_1 = 1\) and the Gaussian noise \(\bm{f}\) is absent, the second equation of (\ref{TT_orig}) can be formally interpreted as a momentum equation in hydrodynamic models \cite{toner1995long}, and one can derive the following energy-dissipation relation:
\begin{equation}
  \begin{aligned}
  \frac{\dd}{\dd t} \int \frac{1}{2} |\bm{v}|^2  \dd x 
  = - \int \left[ (\beta |{\bm v}|^2 -\alpha) |\bm{v}|^2  + D_L |\nabla \cdot \bm{v}|^2 + D_1 |\nabla \bm{v}|^2  \right. \\
   \left. \quad - D_2 \left((\bm{v}\cdot \nabla)^2 \bm{v}\right)\cdot \bm{v} - P(n) (\nabla \cdot \bm{v})  \right]  \dd x.
  \end{aligned}
\end{equation}
Clearly, the right-hand side of this energy law does not exhibit Galilean invariance or satisfy the frame indifference. Moreover, the total energy not always decay with respect to time. These features highlight the nonequilibrium and active nature of the model, distinguishing it from conventional hydrodynamic systems.

Different variants of the TT model have been proposed \cite{gowrishankar2012active, choi2024global}. For instance, \cite{gowrishankar2012active} proposed a model of the form
\begin{equation}\label{PPTT}
  \begin{cases}
    & \pp_t n + \nu_0 \nabla \cdot (n {\bm p}) =  D \Delta n . \\
    & \pp_t {\bm p} + \lambda {\bm p} \cdot \nabla {\bm p} = \alpha {\bm p} - \beta |{\bm p}|^2 {\bm p} + K_1 \nabla (\nabla \cdot {\bm p}) + K_2 \Delta {\bm p} - \xi \nabla c + {\bm f}, \\
  \end{cases}.
\end{equation}
where $n(\x, t)$ is the local concentration, and ${\bm p}$ is the polarization field. The model is simiar to TT model with $P(n) = \xi (n - n_0)$. The main discrepancy lies in the absence of $({\bm v} \cdot \nabla)^2$ in the second equation and the inclusion of a diffusion term in the number density equation. The model is refered as  Parabolic-Parabolic Toner--Tu (PPTT) model in \cite{choi2024global}.

Although the structure of PPTT model resembles that of the original TT model, it provides a different physical intepretation. In the TT model, ${\bm v}$ represents the velocity field of self-propelled particles and simultaneously encodes their polarization direction. In contrast, the PPTT model views ${\bm p}$ as a polarization field. If we define $\uvec = \lambda {\bm p}$ as the background velocity, we can interpret $\pp_t {\bm p} + (\uvec \cdot \nabla ){\bm p}$ as material derivative of the direct field ${\bm p}$ \cite{lin1995nonparabolic}, as the right-hand side is a relaxation dyanmics of ${\bm p}$ (without the noise term), which minmizing a free energy 
\begin{equation}\label{FE_p}
 \mathcal{F}[p] = \int  - \frac{\alpha}{2} |{\bm p}|^2 +  \frac{\beta}{4} |\bm p|^4 +  K_1 (\nabla \cdot {\bm p})^2 + K_2 |\nabla {\bm p}|^2 + \xi \nabla n \cdot {\bm p}
\end{equation}
The free energy (\ref{FE_p}) can be viewed as a special case of the Ericksen-type free energy, commonly used to describe nematic liquid crystals with variable degree of ordering \cite{ericksen1991liquid,lin1995nonparabolic}, with an additional coupling term between the director field and the number density.
Because of this connection, TT-type models can also be interpreted as continuum models for active nematics, even though they are often referred to as models for active polar fluids \cite{shaebani2020computational} due to the use of a vector order parameter that, in principle, lacks head-to-tail symmetry. 

Without loss of generality, in the following, we'll assume $\alpha  = \beta > 0$ and $K_1 = 0$. We also consider the case without the noise term. Under these assumptions, the PPTT model (\ref{PPTT}) reduces to the following simplified form:
\begin{equation}\label{TT_simple}
\begin{cases}
& \pp_t n + \nabla \cdot (\nu_0  {\bm p} n) = \eta \Delta n \\
&   \pp_t {\bm p} + \lambda ({\bm p} \cdot \nabla) {\bm p} =  \gamma (  \Delta {\bm p} - \frac{1}{\epsilon^2} (|{\bm p}|^2  - 1) {\bm p} ) - \zeta \nabla c
\end{cases}
\end{equation}
In the case that the $\nu_0 = \lambda = \zeta = 0$, the system (\ref{TT_simple}) can be viewed as a passive nematic system, which satisfies the energy-dissipation law
\begin{equation}\label{ED_p_c_p}
\frac{\dd}{\dd t} \mathcal{F} (n, {\bm p}) = - \int \eta n |\nabla \ln n|^2 + \frac{1}{\gamma}|\pp_t {\bm p}|^2   \dd \x\ ,
\end{equation}
where the free energy is given by
\begin{equation}
\mathcal{F}(n, {\bm p}) = \int n (\ln n - 1) + \frac{1}{2}|\nabla {\bm p}|^2 +  \frac{1}{4 \epsilon^2} (|{\bm p}|^2 - 1)^2 \dd \x
\end{equation}
The energy-dissipation law (\ref{ED_p_c_p}) can be viewed as a $L^2$-gradient flow with respect to the orientational order parameter ${\bm p}$, and a Wasserstein type gradient flow, or diffusion, with respect to the number density $c$ \cite{giga2017variational}.  There is no coupling between number density $n(\x, t)$ and the orientational order parameter ${\bm p}$. If one further assume that $n$ is constant, then the system reduces to a phenomenological model for passive nematic systems associated with modified Oseen-Frank free energy under the assumption of one constant approximation, as the $ \frac{1}{4\epsilon^2} (|{\bm p}|^2  - 1)^2$ term can be viewed as a penalty term for the unit length constraint \cite{lin1995nonparabolic}.

\begin{rmk}
Similarly, ``dry'' active nematics with a tensor order parameter defined as 
\(\Qvec(\x, t) \) can be written as \cite{narayan2007long,ramaswamy2003active,shaebani2020computational}
\begin{equation}\label{Active_Q}
  \begin{aligned}
    & \partial_t n +  \nabla \cdot \left( \zeta \nabla \cdot \Qvec + J_{\rm passive} \right) = 0, \\
    & \frac{1}{\gamma} \partial_t \Qvec = - \frac{\delta \mathcal{F}[n, \Qvec]}{\delta \Qvec} + {\bm f},
  \end{aligned}
\end{equation}
Here, $\Qvec$ is a symmetric, traceless matrix, and the free energy functional \(\mathcal{F}[n, \Qvec]\) is given by
\begin{equation}
  \mathcal{F}[n, \Qvec] = \int \left[ n \ln n 
  + \left( \frac{A}{2} |\Qvec|^2 
  + \frac{B}{3} \mathrm{tr}(\Qvec^3) 
  + \frac{C}{4} |\Qvec|^4 \right) 
  + \frac{L}{2} |\nabla \Qvec|^2 \right] \, \dd \x,
\end{equation}
which consists of the classical Landau--de Gennes free energy~\cite{de1993physics} and an entropic contribution from the number density. The term \(\zeta \nabla \cdot \Qvec\) represents the active flux in the density evolution. An advantage of using the tensor order parameter \(\Qvec\) is that it naturally preserves the head-to-tail symmetry of nematic molecules.

The mathematical structure of~\eqref{Active_Q} is closely related to the simplified Toner--Tu model~\eqref{TT_simple}. The primary difference lies in the absence of a self-advection term in the evolution equation of \(\Qvec\), which plays a key role in the original Toner--Tu dynamics.

\end{rmk}

Despite their success in capturing key phenomenology of active polar fluids or active nematics, existing Toner--Tu-type models suffer from several important limitations. One major drawback is the absence of a well-defined energy-dissipation law. These models are typically constructed phenomenologically, guided by symmetry considerations and coarse-graining arguments, rather than derived by nonequilibrium-thermodynamics-based variational principles. As a result, they generally lack a clear connection to an underlying free energy functional. This lack of thermodynamic consistency makes it difficult to interpret or control the balance between energy input (due to activity) and dissipation (due to friction, viscosity, or diffusion). 
Consequently, while these TT-type models provide valuable insight into large-scale collective dynamics, they remain limited in their ability to describe energy transduction.

In many biological and synthetic active systems, activity originates from the conversion of chemical energy (e.g., ATP hydrolysis, fuel consumption), which naturally suggests the inclusion of reaction kinetics. However, without a consistent framework, it becomes unclear how to incorporate these reactions systematically due to the two-way coupling between the chemical and mechanical parts. To overcome the limitations of phenomenological models, we aim to derive models of active nematics that incorporate chemo-mechanical coupling within the framework of EnVarA, which has been proved to be a powerful tool to build thermodynamically consistent models for various complex system.

\subsection{Energetic Variational Approach for Chemo-Mechanical Systems}

In this subsection, we briefly introduce the EnVarA, a general framework that offers a unified and thermodynamically consistent methodology for modeling complex systems.

The framework of EnVarA is rooted in the non-equilibrium thermodynamics \cite{de2013non, kjelstrup2008non}, especially the seminal works of Rayleigh \cite{Ra73} and Onsager \cite{On31, On31a}.
The key idea is to describe a complex system by an energy-dissipation law. The energy-dissipation law, together with the kinematics of the employed variables, describe all the physics and assumptions in the system. The dynamics and the constitutive equation can be derived through variational procedures. 
In more details, an energy-dissipation law (for an isothermal closed system) can be written as
\begin{equation}\label{ED}
\frac{\dd}{\dd t} E^{\rm total} = - \triangle(t),
\end{equation}
where $E^{\rm total}$ is the total energy, including both the kinetic energy $\mathcal{K}$ and the Helmholtz free energy $\mathcal{F}$, and $\triangle(t) \geq 0$ is the rate of the energy dissipation which is equal to the entropy production in the isothermal case. 

The existence of energy-dissipation law \eqref{ED} follows directly from the first and second laws of thermodynamics \cite{ericksen1992introduction}. The first law expresses conservation of energy as
\begin{equation}\label{1stlaw}
\frac{\dd}{\dd t} (\mathcal{K} + \mathcal{U}) = \delta W + \delta Q,
\end{equation}
where $\mathcal{K}$ and $\mathcal{U}$ denote the kinetic and internal energy, respectively, and $\delta W$, $\delta Q$ represent the rate of mechanical work and heat input. These quantities are path-dependent and are not exact differentials.
To quantify the heat exchange, one needs to consider the second law of thermodynamics, which states
\begin{equation}\label{2ndlaw}
T \dd S = \delta Q + \triangle,
\end{equation}
where $T$ is the absolute temperature, $S$ is the entropy, and $\triangle \geq 0$ denotes the rate of entropy production. For an isothermal and mechanically isolated system (i.e., $\delta W = 0$ and $T$ constant), subtracting \eqref{2ndlaw} from \eqref{1stlaw} yields the energy-dissipation law \eqref{ED}, as the free energy $\mathcal{F} = \mathcal{U} - T S$.
For non-isothermal systems, the situation becomes much more complicated. We refer interested readers to \cite{Jan-Eric} and \cite{wang2022some} for detailed treatments.

Starting with an energy-dissipation law, EnVarA derives the dynamics of the systems through the Least Action Principle (LAP) and the Maximum Dissipation Principle (MDP). The LAP, which states the equation of motion for a Hamiltonian system
can be derived from the variation of the action functional $\mathcal{A} = \int_{0}^T \mathcal{K} - \mathcal{F} \dd t$ with respect to the flow map $\x(\X, t)$, gives a unique procedure to derive the conservative force for the system.
For fixed $\X$, the flow map $\x(\X, t)$ can be viewed as the trajectory of a particle initially at position $X$. For fixed $t$, the flow map $\x(\X, t)$ defines a diffeomorphism from a reference domain $\Omega_0$ to the current domain $\Omega$.
The MDP, variation of the dissipation potential $\mathcal{D}$, which equals to $\frac{1}{2}\triangle$ in the linear response regime, with respect to $\x_t$, i.e., the velocity, gives the dissipation force for the system. In turn, the force balance condition leads to the evolution equation of the system
\begin{equation}\label{EnVarA_1}
\frac{\delta \mathcal{D}}{\delta \x_t} = \frac{\delta \mathcal{A}}{\delta \x},
\end{equation}
The variational procedure guarantee the resulting evolution equation is thermodynamically consistent, which is crucial for establishing well-posedness of the resulting system. 

As an example, we consider how to model a generalized diffusions using EnVarA. A generalized diffusion is concerned with a conserved quantity $\rho(\x, t)$ satisfying the kinematics, i.e., the conservation of mass,
\begin{equation}\label{kinematics_1}
\pp_t \rho + \nabla \cdot (\rho \uvec) = 0,
\end{equation}
where $\uvec$ is the average velocity. 
The energy-dissipation law of a generalized diffusion can be written as
\begin{equation}
\frac{\dd}{\dd t} \mathcal{F}[\rho] = - \int \eta(\rho) |\uvec|^2 \,\dd \x, \quad  \mathcal{F}[\rho]  = \int \omega(\rho) \,\dd \x,
\end{equation}
where $\omega(\rho)$ is the free energy density, $\eta(\rho)$ is the friction coefficient. We emphasize that throughout this paper, we always consider a non-dimensionalized setting.

Due to the kinematics (\ref{kinematics_1}), the free energy can be reformulated as a functional of $\x(\X, t)$ in Lagrangian coordinates.
A direct computation shows that
\begin{equation*}
  \begin{aligned}
    & \delta \mathcal{A} = - \delta \int_{0}^T  \int_{\Omega_0} \omega(\rho_0(X)/ \det F) \det F \, \dd \X \dd t \\
    & = - \int_{0}^T \int_{\Omega_0} \left( - \frac{\pp \omega}{\pp \rho}  \left( \frac{\rho_0(X)}{\det F} \right)  \cdot \frac{\rho_0(X)}{\det F} + \omega\left(\frac{\rho_0(X)}{\det F}\right) \right)  \times  (F^{-\rm{T}} : \nabla_{\X} \delta \x)\det F \, \, \dd \X \dd t, \\
      \end{aligned}
\end{equation*}
where $\delta \x(\X, t)$ is the test function satisfying $\delta \x \cdot \n = 0$ with $\n$ is the outer normal of $\Omega$ in Eulerian coordinates (Here we will not distinguish $\tilde{\delta \x}(\x(\X, t), t) = \delta \x(\X, t)$ and $\delta \x(\X, t)$ without ambiguity). Push forward to Eulerian coordinates, we have
\begin{equation}\label{LAP1}
  \begin{aligned}
\delta \mathcal{A} & = - \int_{0}^T \int_{\Omega} ( -  \frac{\pp \omega}{\pp \rho} \rho + \omega) \nabla \cdot (\delta \x) \dd \x = \int_{0}^T \int_{\Omega} - \nabla  (\frac{\pp \omega}{\pp \rho} \rho - \omega) \cdot \delta \x  \dd \x \dd t, \\
  \end{aligned} 
\end{equation}
which indicates that $$\frac{\delta \mathcal{A}}{\delta \x} = - \nabla (\frac{\pp \omega}{\pp \rho} \rho - \omega) = - \rho \nabla \mu,$$ where $\mu = \frac{\delta \mathcal{F}}{\delta \rho}$ is the chemical potential. In the notion of the principle of virtual work \cite{berdichevsky_variational_2009}, one can obtained (\ref{LAP1}) by using the relation $\delta \rho = \nabla \cdot (\rho \delta \x)$. For the dissipation part, since $\mathcal{D} = \frac{1}{2} \int  \eta(\rho) |\uvec|^2  \dd \x$ it is easy to compute that $\frac{\delta \mathcal{D}}{\delta \uvec} = \eta(\rho) \uvec$. As a consequence, we have the force balance equation
\begin{equation}\label{FB1}
  \eta(\rho) \uvec =  - \rho \,\nabla \mu.
\end{equation}
Combining the force balance equation (\ref{FB1}) with the kinematics (\ref{kinematics_1}), one obtain a generalized diffusion equation
\begin{equation}\label{G_diffusion}
\pp_t \rho = \nabla \cdot \left( \frac{\rho^2}{\eta(\rho)}   \nabla \mu \right).
\end{equation}

\begin{rmk}
For the $L^2$-gradient flow
\begin{equation}
  \frac{\dd}{\dd t} \mathcal{F}[\varphi] \dd x = - \int_{\Omega} \frac{1}{\gamma} |\pp_t \varphi|^2 \dd \x\ ,
\end{equation}
One may apply the EnVarA by viewed $\varphi$ as generalized coordinates \cite{doi2011onsager}, which leads to  
\begin{equation}\label{L2_gradient_flow}
\frac{1}{\gamma} \pp_t \varphi = - \frac{\delta \mathcal{F}}{\delta \varphi}
\end{equation}
Alternatively, one can impose the kinematics $\pp_t \varphi  + \uvec \cdot \nabla \varphi = 0$ \cite{liu2020variational} and rewrite the energy-dissipation law as
\begin{equation}
\frac{\dd}{\dd t} \mathcal{F}[\varphi] \dd x = - \int_{\Omega} \frac{1}{\gamma} |\uvec \cdot \nabla \varphi|^2 \dd \x\ ,
\end{equation}
By performing the EnVarA with respect to the flow map $\x(\X, t)$ and the velocity $\uvec(\x, t)$, one will end up with, at least formally,
\begin{equation}
 \frac{1}{\gamma} (\nabla \varphi \cdot \uvec) \nabla \varphi  =  \frac{\delta \mathcal{F}}{\delta \varphi} \nabla \varphi\ ,
\end{equation}
which is equivalent to (\ref{L2_gradient_flow}) when $\nabla \varphi \neq 0$.

\end{rmk}

Classical EnVarA approach is developed for mechanical systems, and the variation are taken with respect to the flow map $\x(\X, t)$ and its time derivative, i.e., the velocity $\uvec(\x, t)$.
In \cite{wang2020field}, we formulate a reaction kinetics with detailed balance in a energetic variational form by using the reaction trajectory ${\bm R}$ \cite{oster1974chemical}, which is an analogy to the flow map in mechanical system. Roughly speaking, the reaction trajectory ${\bm R}$ accounts for the number of forward reaction has occurred by time $t$. The reaction rate ${\bm r}$ is defined as $\pp_t {\bm R}$, which is the reaction velocity \citep{kondepudi2014modern}. The reaction trajectory is known as extent of reaction in physical chemistry, and was introduced by de
Donder in 1920s \cite{de1920leccons}.

For a general reversible chemical reaction system containing $N$ species $\{ X_1, X_2, \ldots X_N \}$ and $M$ reactions can be represented by 
$$\ce{ \alpha_{1}^{l} X_1 + \alpha_{2}^{l}X_2 + \ldots \alpha_{N}^{l} X_N <=>[k_l^+][k_l^-] \beta_{1}^{l} X_1 + \beta_{2}^{l}X_2 + \ldots \beta_{N}^{l} X_N}, \quad l = 1, \ldots, M.$$
The relation between species concentration ${\bm c} \in \mathbb{R}_{+}^N$ and the reaction trajectory ${\bm R} \in \mathbb{R}^M$ is given by ${\bm c} = {\bm c}_0 + {\bm \gamma} {\bm R},$ where ${\bm c}_0$ is the initial concentration, and $\bm{\gamma} \in \mathbb{R}^{N \times M}$ is the stoichiometric matrix with $\gamma_{il} = \beta^l_i - \alpha^l_i$. Within the reaction trajectory, we can describe the chemical kinetics by energy-dissipation law in terms of ${\bm R}$ and ${\bm R}_t$:
\begin{equation}
  \frac{\dd}{\dd t} \mathcal{F}_{\rm chem}[{\bm R}] = - \triangle_{\rm chem}[{\bm R}, {\bm R}_t],
  \end{equation}
  where $\triangle_{\rm chem}[{\bm R}, {\bm R}_t]$ is the rate of energy dissipation due to  the chemical reaction procedure. Since the linear response assumption may not be valid for chemical reactions \citep{beris1994thermodynamics, de2013non}, it is often assumed that $\mathcal{D}_{\rm chem}[{\bm R}, {\bm R}_t] = \left( {\bm \Gamma}({\bm R}, {\bm R}_t),  {\bm R}_t  \right) = \sum_{l=1}^M \Gamma_l ({\bm R},  {\bm R}_t) \pp_t R_l \geq 0,$ which leads to the reaction rate can be derived as \citep{wang2020field,liu2021structure}:
  \begin{equation}\label{eq_Rl}
  \Gamma_l({\bm R}, {\bm R}_t) = - \frac{\delta \mathcal{F}_{\rm chem}}{\delta R_l},
  \end{equation}
  In this formulation, the choice of the free energy determines the chemical equilibrium, while the choice of the dissipation functional $\mathcal{D}_{\rm chem}[{\bm R}, \pp_t {\bm R}]$ determines the reaction rate. 

As an example, we consider a simple chemical reaction:
\begin{equation}\label{reaction_example}
  \alpha_1 X_1 + \alpha_2 X_2 \ce{<=>[k^+][k^-]} \beta_3 X_3,
\end{equation}
where $\alpha_i$ and $\beta_i$ are the stoichiometric coefficients. Let $c_i$ denote the concentration of species $X_i$. According to the law of mass action (LMA), the reaction rate is given by
\begin{equation}\label{LMA}
  r = k^+ c_1^{\alpha_1} c_2^{\alpha_2} - k^- c_3^{\beta_3}.
\end{equation}
This system satisfies the detailed balance condition, meaning there exists an equilibrium state $(c_1^\infty, c_2^\infty, c_3^\infty)$ with all $c_i^\infty > 0$ such that
\begin{equation}
  k^+ (c_1^\infty)^{\alpha_1} (c_2^\infty)^{\alpha_2} = k^- (c_3^\infty)^{\beta_3}.
\end{equation}
A corresponding free energy functional can be defined as \cite{mielke2011gradient}
\begin{equation}\label{FE_chem1}
  \mathcal{F}_{\rm chem}[\bm{c}] = \int_{\Omega} \sum_{i=1}^3 c_i \left( \ln \left( \frac{c_i}{c_i^\infty} \right) - 1 \right) \, \dd \x = \int_{\Omega} \sum_{i=1}^3 \left( c_i (\ln c_i - 1) + c_i \sigma_i \right) \, \dd \x,
\end{equation}
where $\sigma_i = -\ln c_i^\infty$ can be interpreted as an internal energy contribution for species $X_i$, reflecting its chemical activity. Clearly, the free energy determines 
\begin{equation}\label{Eq_chem}
 \frac{k^+}{k^-} = \frac{(c_3^\infty)^{\beta_3}}{(c_1^\infty)^{\alpha_1} (c_2^\infty)^{\alpha_2}} .
\end{equation}

We introduce the stoichiometric vector ${\bm \gamma} = ( - \alpha_1, - \alpha_2, \beta_3)$. Then, $c_i(t) = c_i(0) + \gamma_i R(t)$ with $R(t)$ being the reaction trajectory. We can view the free energy (\ref{FE_chem1}) as the functional of $R(t)$.
To determine the dynamics, we impose the dissipation as
\begin{equation}
  \triangle_{\rm chem} = \Gamma(R, R_t) R_t\ ,
\end{equation}
where $\Gamma(R, R_t)$ is chosen such that $\Gamma(R, R_t) R_t \geq 0$. Then we have
\begin{equation}\label{FB_chem}
   \Gamma(R, R_t) =  - \frac{\delta \mathcal{F}_{\rm chem}}{\delta R} = -  \sum_{i=1}^3 \gamma_i \mu_i, \quad \mu_i = \frac{\delta \mathcal{F}_{\rm chem}}{\delta c_i} =  \ln c_i + \sigma_i\ .
\end{equation}
Here $ - \gamma_i \mu_i$ is known as the chemical affinity, which is the driven force in chemical reaction \cite{kondepudi2014modern}. One can interpret (\ref{FB_chem}) as the force balance equation for the chemical part of the system. To derive the classical law of mass action (\ref{LMA}), one can choose 
\begin{equation}
   \Gamma(R, R_t) = \ln \left(  \frac{R_t}{k^- c_3^{\beta_3}} + 1 \right)\ ,
\end{equation}
which leads to $R_t = k^- c_3^{\beta_3} (\exp( - \gamma_i \mu_i) -  1) = k^+ c_1^{\alpha_1} c_2^{\alpha_2} - k^- c_3^{\beta_3}$ by using (\ref{Eq_chem}). Other choice of $\Gamma(R, R_t)$ may lead to different reaction kinetics \cite{wang2020field}.

The energetic variational formulation of chemical reactions opens a new door to model a general chemo-mechanical system in a unified way. For example, we can add the diffusion, a mechanical process, to (\ref{reaction_example}) by assuming $c_i$ satisfies the kinematics: 
  \begin{equation}\label{Kin_1}
   \pp_t c_i + \nabla \cdot (c_i \uvec_i) = \gamma_i R_t, \quad i = 1, 2, \ldots 3 , 
  \end{equation}
  where $\uvec_i(\x, t)$ is the average velocity of each species due to its own diffusion, and $R(\x, t)$ is the reaction trajectory at each space location $\x$. Here, $R_t$ is the partial derivative of $R(x, t)$ with respect to $t$. 
  Then the reaction-diffusion equation can be modeled through the energy-dissipation law \citep{wang2020field}: 
  \begin{equation}\label{ED_RD}
      \frac{\dd}{\dd t} \mathcal{F}_{\rm chem}[{\bm c}]  =  - \int_{\Omega} \left[ \sum_{i=1}^3 \eta_i(c_i) |\uvec_i|^2 +  \Gamma(R, R_t )  R_t  \right] \dd \x. 
  \end{equation}
  We can employ EnVarA to obtain equations for the reaction and diffusion part respectively, i.e., to obtain the ``force balance equation'' of the chemical and mechanical subsystems, which leads to
  \begin{equation}\label{RD_FB}
    \begin{cases}
      & \eta_i(c_i) \uvec_i =  - c_i \nabla \mu_i, \quad i = 1, 2, \ldots 3, \\
      &  \Gamma(R, \pp_t R) = - \sum_{i=1}^3 \gamma_i \mu_i \\
    \end{cases}
  \end{equation}
  By taking $\eta_i(c_i) = \frac{1}{D_i}c_i$, we have a reaction-diffusion system
  \begin{equation}
  \pp_t c_i = D_i \Delta c_i + \gamma_i R_t \ .
  \end{equation}
  More mechanical effects can be included by modifying the energy and the dissipation.

\subsection{Energy-dissipation analysis on a simplified Toner--Tu model}
A natural question is whether the Toner--Tu type models can be derived using the energetic variational approach. To this end, we first perform an energy-dissipation analysis on a Toner--Tu model with the following form
\begin{equation}\label{TT_model_1}
  \begin{cases}
    & \pp_t n + \nabla \cdot ( \nu_0 {\bm p} n) =  \eta \nabla \cdot (n \nabla \mu_{n} ), \quad \mu_n  = \frac{\delta \mathcal{F}}{\delta n} \\
    & \pp_t {\bm p} + \lambda ({\bm p} \cdot \nabla) {\bm p} = - \gamma {\mu_{\bm p}}, \quad {\bm \mu}_{\bm p} = \frac{\delta \mathcal{F}}{\delta {\bm p}}
  \end{cases}
\end{equation}
where the free energy $\mathcal{F}[n, {\bm p}]$ is given by
\begin{equation}\label{TT_model_free_energy}
  \mathcal{F}[n, {\bm p}] = \int n \ln n + \frac{1}{2} |\nabla {\bm p}|^2 + \frac{1}{4 \epsilon^2} (|{\bm p}|^2 - 1)^2 + \xi \nabla c \cdot {\bm p}  \dd \x.
\end{equation}
This formulation corresponds to the simplified Toner--Tu model \eqref{TT_simple}, with an additional term $ - \eta  \nabla \cdot( n \nabla (\xi \nabla \cdot {\bm p}) )$ in the number density equation to ensure consistency with the variational structure of the free energy. Since $\nabla \cdot {\bm p}$ is often small due to the choice of the free energy, it is reasonable to neglect this term in practice.

By a direct calculation, we have the following energy-dissipation relation
\begin{equation}\label{ED_TT}
\begin{aligned}
\frac{\dd}{\dd t} \mathcal{F}[n, {\bm p}] & = \int \left( \mu_n \partial_t n + \mu_{\bm p} \cdot \partial_t \bm{p} \right) \dd \x \\
& = \int \mu_n \left[ - \nabla \cdot (\nu_0 n \bm{p}) + \eta \nabla \cdot (n \nabla \mu_n) \right]  +  \mu_{\bm p} \cdot \left[ -\lambda ({\bm p} \cdot \nabla) \bm{p} - \gamma \mu_{\bm p} \right] \dd \x \\
& = \int - \eta n |\nabla \mu_n|^2  - \gamma |\mu_{\bm p}|^2 + \nu_0 n \bm{p} \cdot \nabla \mu_n - \lambda \mu_{\bm p} \cdot ( ({\bm p} \cdot \nabla) \bm{p} ) \dd \x. \\
\end{aligned}
\end{equation}
Clearly, the system is no longer strictly dissipative if $\nu_0 \neq 0$ and $\lambda \neq 0$, due to the appearance of two active terms arising from the advection of the density and the convective derivative in the polarization equation. The energy-dissipation relation (\ref{ED_TT}) can be written as
\begin{equation}\label{ED_TT_2}
  \begin{aligned}
\frac{\dd}{\dd t} \mathcal{F} (n, {\bm p}) = - & \int \left( n |\uvec - \nu_0 {\bm p}|^2 + \frac{1}{\gamma}|\pp_t {\bm p}  + \lambda ({\bm p} \cdot \nabla) {\bm p}|^2  \right) \dd \x \\ 
& + \int \nu_0 n \bm{p} \cdot \nabla \mu_n - \lambda \mu_{\bm p} \cdot ( ({\bm p} \cdot \nabla) \bm{p} )  \dd \x.  \\
  \end{aligned}
\end{equation}
where $\uvec$ is the transport velocity for the number density such that the number density satisfies the kinematics $\pp_t n  + \nabla \cdot (n \uvec) = 0$.  However, one cannot apply the energetic variational approach to (\ref{ED_TT_2}) to derive the underlying equation.

\begin{rmk}
  Formally, one may start with the energy-dissipation law
  \begin{equation}\label{ED_TT_11}
    \frac{\dd}{\dd t} \mathcal{F}(n, {\bm p}) = - \int \left( n |\uvec - \nu_0 {\bm p}|^2 + \frac{1}{\gamma} \left| \partial_t {\bm p} + \lambda ({\bm p} \cdot \nabla) {\bm p} \right|^2 \right) \dd \x,
  \end{equation}
  and apply the energetic variational approach to derive the force balance equations
  \begin{equation}\label{FB_TT}
    \begin{aligned}
      & n (\uvec - \nu_0 {\bm p}) = - n \nabla \mu_n, \\
      & \frac{1}{\gamma} \partial_t {\bm p} + \lambda ({\bm p} \cdot \nabla) {\bm p} = - {\bm \mu}_{\bm p}.
    \end{aligned}
  \end{equation}
  However, as shown in the earlier computation (\ref{ED_TT}), the system (\ref{FB_TT}) does not yield the original energy-dissipation law (\ref{ED_TT_11}). The discrepancy arises because ${\bm p}$ is not a true velocity field. Consequently, $\uvec - \nu_0 {\bm p}$ cannot be interpreted as a relative drag velocity, and the term $\partial_t {\bm p} + \lambda ({\bm p} \cdot \nabla) {\bm p}$ does not represent a material transport as in classical liquid crystal models \cite{lin1995nonparabolic}. In other words, the presence of active transport terms destroys the variational structure, which is preserved only in the case $\nu_0 = \lambda = 0$.

  To address this issue, one possible strategy is to introduce an auxiliary velocity field ${\bm V}$ representing the macroscopic flow, and consider the modified energy-dissipation law
  \begin{equation}
    \frac{\dd}{\dd t} \mathcal{F}(n, {\bm p}) = - \int \left( n |\uvec - \kappa {\bm V}|^2 + \frac{1}{\gamma} \left| \partial_t {\bm p} + ({\bm V} \cdot \nabla) {\bm p} \right|^2 \right) \dd \x,
  \end{equation}
  together with the constraint ${\bm V} = \lambda {\bm p}$. This constraint can be relaxed, for example, by enforcing ${\bm V} \times {\bm p}$ to remain small. Nevertheless, it remains challenging to incorporate this condition within a thermodynamically consistent variational framework and to derive a well-posed system involving both ${\bm V}$ and ${\bm p}$. We leave a detailed investigation of this direction for future work.

  This limitation motivates the development of alternative strategies for uncovering the variational structure of Toner--Tu-type models by explicitly incorporate the chemical reaction that powers active transport into the model.
\end{rmk}

\section{Toner--Tu Model with ATP hydrolysis}

In the previous section, we discussed the challenges in formulating a variational structure for Toner--Tu type models, primarily due to the presence of active transport terms. Without explicitly specifying the energy source driving the activity, the system behaves as an open system lacking a well-defined energy-dissipation law, and thus cannot be derived within the energetic variational framework.

In this section, we show that by explicitly incorporating the chemical reaction for the originality of the activity into the model, we can have a thermodynamically consistent dissipative model. The simplified Toner--Tu model (\ref{TT_model_1}) can be obtained by setting the reaction rate as a constant.
This coupling not only quantifies the energetic cost of activity but also ensures that the system becomes passive in the absence of ATP consumption.

\subsection{Model derivation}

To explain the idea, we assume that the primary energy source for the activity comes from the ATP hydrolysis 
\begin{equation}\label{ATP_reaction}
  \ce{ATP <=> ADP + P}.
\end{equation}
The free-energy change for the hydrolysis of
each ATP molecule is given by
\begin{equation}
 \triangle \mu = \mu_{\rm ATP} - \mu_{\rm ADP} - \mu_{\rm P},
\end{equation}
which indeed in the chemical affinity of ATP hydrolysis \cite{wang2021onsager}. Here, $\mu_{\rm ATP}$, $\mu_{\rm ADP}$, and $\mu_{\rm P}$ are the chemical potentials of ATP, ADP, and P, respectively, which is associated with a free energy
\begin{equation}\label{free_energy_ATP}
\mathcal{F}_{\rm chem}({\bm c}) = \int \sum_{i=1}^3 c_i (\ln c_i - 1) + c_i \sigma_i \dd \x.
\end{equation}
where $c_1$, $c_2$ and $c_3$ are the concentrations of ATP, ADP and P, respectively, and $\sigma_i$ represent the internal energy associated with each chemical species, which contribute to the activity of each species. Following the general framework of modeling reaction-diffsuion equation, we assumm $c_i$ satisfies the kinematics
\begin{equation}
\pp_t c_i  + \nabla \cdot (c_i \uvec_i) = \gamma_i R_t, \quad i = 1, 2, 3,
\end{equation}
where $\uvec_i$ is the transport velocity of $c_i$, $R(x, t)$ is the reaction trajectory of the ATP hydrolysis (\ref{ATP_reaction}). The stoichiometric vector $(\gamma_1, \gamma_2, \gamma_3) = (-1, 1, 1)$ in this case and $\triangle \mu = - \sum_{i=1}^3 \gamma_i \mu_i$.

The key idea of coupling the ATP hydrolysis with the Toner--Tu model (\ref{TT_model_1}) is to asssume that the coefficients in the active transprot, i.e., $\mu_0$ and $\lambda$ in (\ref{TT_model_1}), depend on the reation rate of ATP hydrolysis. Both coefficients go to zero when the ATP hydrolysis stops, i.e., the reaction rate is zero. We can derive the overall model from the energy-dissipation law
\begin{equation}\label{eq:total_energy_balance}
\begin{aligned}
\frac{\dd}{\dd t} \left(\mathcal{F}(n, \bm{p}) + \mathcal{F}_{\text{chem}}({\bm c})\right)  = & - \int \left( \frac{n}{\eta} |\bm{u} - \nu_0 R_t \bm{p}|^2 + \frac{1}{\gamma} |\partial_t \bm{p} + \lambda R_t (\bm{p} \cdot \nabla) \bm{p}|^2 \right. \\
   & \left.  \quad \quad  \quad \quad  +  \sum_{i=1}^3 \frac{c_i}{\eta_i} |\uvec_i|^2 + \Gamma(R, R_t) R_t \right) \dd \x  \\ .
\end{aligned}
\end{equation}
where $\mathcal{F}(n, {\bm p})$ is the mechanical free energy defined in (\ref{TT_model_free_energy}). 


By perfoming the energetic variational approach to each componets, we have
\begin{equation}
  \begin{aligned}  
    &  n (\bm{u} - \nu_0 R_t \bm{p}) = -  \eta n \nabla \mu_n, \quad \mu_n  = \frac{\delta \mathcal{F}}{\delta n}   \\ 
    &  \partial_t \bm{p} + \lambda R_t (\bm{p} \cdot \nabla) \bm{p} = - \gamma {\bm \mu}_p, \quad \mu_{\bm{p}} = \frac{\delta \mathcal{F}}{\delta {\bm p}}  \\
    &  c_i \uvec_i = - \eta_i \nabla \mu_i, \quad \mu_i = \frac{\delta \mathcal{F}_{\rm chem}}{\delta c_i} \\ 
    &    \Gamma(R, R_t) - \frac{n}{\eta} (\bm{u} - \nu_0 R_t \bm{p}) \cdot (\nu_0 {\bm p}) + \frac{1}{\gamma} (\partial_t \bm{p} + \lambda R_t (\bm{p} \cdot \nabla) {\bm p}) \cdot (\lambda (\bm{p} \cdot \nabla) {\bm p}) = - \sum_{i=1}^3 \gamma_i \mu_i
  \end{aligned}
\end{equation}

The reaction rate equation can be written as
\begin{equation}\label{eq_R_ATP}
\Gamma(R, R_t) =  - \sum_{i=1}^3 \gamma_i \mu_i - n \nabla \mu_n \cdot (\nu_0 p) +  \lambda {\bm \mu}_{\bm p} \cdot (\bm{p} \cdot \nabla) {\bm p}\ ,
\end{equation}
which can be reduced to the reaction rate equation (\ref{FB_chem}) if $\lambda = \nu_0 = 0$. The equation (\ref{eq_R_ATP}) reveals the chemo-mechanical coupling, in which the reaction rate is modulated by the mechanical process. One can choose $$\Gamma(R, R_t) = \ln \left( \frac{R_t}{k^{-} c_2 c_3} + 1 \right)\ ,$$ which recovers the law of mass action in the absence of mechanical coupling (i.e. $\lambda = \nu_0 = 0$). But in the current study, we take $$\Gamma(R,R_t) = \Lambda R_t$$ with $\Lambda > 0$, i.e., the linear response assumption, for simplicity. 
Then the final dynamics are governed by the following system:
\begin{equation}\label{Final_eq}
\begin{cases}
\pp_t n + \nabla \cdot ( \nu_0 R_t {\bm p} n) =  \eta \nabla \cdot (n \nabla \mu_{n} ), \quad \mu_n  = \frac{\delta \mathcal{F}}{\delta n} \\ 
\partial_t \bm{p} + \lambda R_t (\bm{p} \cdot \nabla) \bm{p} = - \gamma \mu_{\bm{p}}, \mu_{\bm{p}} = \frac{\delta \mathcal{F}}{\delta \bm{p}}, \\
\pp_t c_i = \eta_i  \nabla \cdot ( c_i \nabla \mu_u) + \gamma_i R_t   \\
\Lambda R_t = \triangle \mu - \nu_0 n (\bm{u} \cdot \bm{p}) + \lambda \mu_{\bm{p}} \cdot (\bm{p} \cdot \nabla) \bm{p},
\end{cases}
\end{equation}
where $\triangle \mu = \mu_{\rm ATP} - \mu_{\rm ADP} - \mu_{\rm P}$ is the chemical affinity. The model is reduced to the simplified Toner--Tu model (\ref{TT_model_1}) if $R_t$ is a constant. 

It is worth highlighting that the final system (\ref{Final_eq}) exhibits a two-way chemo-mechanical coupling. On the one hand, the active transport in the mechanical subsystem is modulated by the chemical reaction rate. On the other hand, the reaction rate itself is influenced by the mechanical interactions. A key advantage of the energetic variational approach is that it ensures this bidirectional coupling remains thermodynamically consistent by construction.

\begin{rmk}
Unlike the issue that we mentioned in remark 2.2, we can prove the final equation (\ref{Final_eq}) satisfies the energy-dissipation law (\ref{eq:total_energy_balance}), since ${\bm p}R_t$ becomes a cetrain ``velocity'', induced by the chemical reaction. Indeed, by a direct calculation, we have
  \begin{equation}
    \begin{aligned}
      & \frac{\dd}{\dd t} \left( \mathcal{F} [n, {\bm p}] + \mathcal{F}_{\rm chem} [{\bm c}] \right) = \int \mu_n \pp_t n  + {\bm \mu}_p \cdot \pp_t  {\bm p} +  \sum_{i=1}^3 \mu_i  \pp_t c_i  ~ \dd x  \\
      & = \int \mu_n (\eta \nabla \cdot (n \nabla \mu_{n} ) -   \nabla \cdot ( \nu_0 R_t {\bm p} n) )  +  {\bm \mu}_p \cdot (  - \gamma \mu_{\bm{p}} - \lambda R_t (\bm{p} \cdot \nabla) \bm{p} )  \\
      & \quad \quad  +  \sum_{i=1}^3 \mu_i ( \eta_i  \nabla \cdot ( c_i \nabla \mu_u) + \gamma_i R_t )  \dd x \\
      & = \int - \eta n |\nabla \mu_n|^2 - \gamma |\mu_p|^2 - \eta_i c_i |\nabla \mu_i|^2 + R_t ( \sum_{i=1}^3 \gamma_i \mu_i + \nu_0 n \nabla \mu_n \cdot {\bm p} - \lambda {\bm \mu}_{\bm p} \cdot ({\bm p} \cdot \nabla {\bm p}) ) \\
      & = \int - \eta n |\nabla \mu_n|^2 - \gamma |\mu_p|^2 - \eta_i c_i |\nabla \mu_i|^2  - \Lambda |R_t|^2 \dd x \leq 0
    \end{aligned}
  \end{equation}
\end{rmk}

\begin{rmk}
It is straightforward to apply the same framework to revise the ``dry'' active nematic models with $\Qvec$-tensor \cite{narayan2007long} using the energy-dissipation law
  \begin{equation}
  \frac{\dd}{\dd t} \mathcal{F}[\Qvec, n] + \mathcal{F}_{\rm chem} [{\bm c}] = - \int c |\uvec -  \zeta R_t \nabla \cdot \Qvec|^2 + \frac{1}{\gamma}  |\pp_t \Qvec|^2 + \sum_{i=1}^n \frac{c_i}{\eta_i} |\uvec_i|^2 + \Gamma(R, R_t) R_t \dd \x \ ,
\end{equation}
where $n$ is the number density of the liquid crystal molecules, ${\Qvec}$ is the tensor order parameter, and $\mathcal{F}[\Qvec, n]$ is the free energy for the mechanical part, and $\mathcal{F}_{\rm chem} [{\bm c}]$ is the free energy associated with the ATP hydrolysis procedure, defined in (\ref{free_energy_ATP}). We'll study this model in future work.
\end{rmk}

\subsection{Energy transduction and efficiency}

A key feature of the proposed model is its ability to quantify the energy transduction, the free energy released by ATP hydrolysis into the mechanical activity. This is achieved through the coupling of the reaction coordinate \( R_t \) with the active transport coefficient in the number density and the polarization equations.

The total free energy of the system consists of two parts: the mechanical free energy \(\mathcal{F}(n, \bm{p})\), and the chemical free energy \(\mathcal{F}_{\mathrm{chem}}({\bm c})\). The previous analysis shows that the term
\begin{equation}
R_t (\nu_0 n {\bm p} \cdot \nabla \mu_n  - \lambda {\bm \mu}_p \cdot ( ({\bm p} \cdot \nabla ) {\bm p}))
\end{equation}
may contribute to the increase of the mechanical energy \(\mathcal{F}(n, \bm{p})\). Hence, the energy transduction happend when
\begin{equation}
\nu_0 n {\bm p} \cdot \nabla \mu_n  - \lambda {\bm \mu}_p \cdot ( ({\bm p} \cdot \nabla ) {\bm p}) > 0, \quad R_t > 0
\end{equation}
We can define the local chemo-mechanical transduction efficiency as
\begin{equation}
  \eta_{\rm loc}(\x, t) = \frac{ \nu_0 n {\bm p} \cdot \nabla \mu_n - \lambda {\bm \mu}_{\bm p} \cdot (({\bm p} \cdot \nabla) {\bm p}) }{ \Delta \mu },
\end{equation}
where $\Delta \mu$ is the chemical affinity associated with ATP hydrolysis. This dimensionless quantity measures the proportion of chemical energy that is effectively transduced into mechanical contributions at each spatial location. For a global efficiency measure, one may define
\begin{equation}
  \eta_{\mathrm{tot}}(t) = \frac{ \int R_t \left( \nu_0 n {\bm p} \cdot \nabla \mu_n - \lambda {\bm \mu}_{\bm p} \cdot (({\bm p} \cdot \nabla) {\bm p}) \right) \dd \x }{ \int R_t \Delta \mu \, \dd \x },
\end{equation}
which quantifies the net transduction efficiency over the entire domain.

It is important to note that even when the transduction efficiency is positive, the mechanical energy $\mathcal{F}(n, {\bm p})$ may not increase in time, due to dissipation in both the $n$ and ${\bm p}$ dynamics. The increasing of free energy only occurs when the chemical driving overcomes the dissipative losses in the system.

\begin{rmk}
  Without chosen model parameters properly, one may have $R_t < 0$ or $\nu_0 n {\bm p} \cdot \nabla \mu_n  - \lambda {\bm \mu}_p \cdot ( ({\bm p} \cdot \nabla ) {\bm p}) < 0$, which may lead to an increase in the chemical energy $\mathcal{F}_{\rm chem}$, indicating that mechanical energy is being converted back into chemical energy. Although this scenario is non-physical in the current setting, it highlights the importance of parameter selection to ensure that the model reflects the correct direction of energy transduction.

\end{rmk}

\section{Numerics}
In this section, we provide some numerical simulation result to new Toner--Tu model with ATP hydrolysis.
\begin{equation}
\begin{cases}
\pp_t n + \nabla \cdot ( \nu_0 R_t {\bm p} n) =  \eta \nabla \cdot (n \nabla \mu_{n} ), \quad \mu_n  = \ln n - \xi (\nabla \cdot {\bm p}) \\ 
\partial_t \bm{p} + \lambda R_t (\bm{p} \cdot \nabla) \bm{p} = - \gamma \mu_{\bm{p}}, \mu_{\bm{p}} = - \Delta {\bm p} + \frac{1}{\epsilon^2} (|{\bm p}|^2 - 1) {\bm p} + \xi \nabla c, \\
\pp_t c_i = \eta_i  \nabla \cdot ( c_i \nabla \mu_u) + \gamma_i R_t   \\
\Lambda R_t = \triangle \mu - \nu_0 n ( \nabla \mu_n \cdot \bm{p}) + \lambda \mu_{\bm{p}} \cdot (\bm{p} \cdot \nabla) \bm{p}, \\
\triangle \mu = \ln \left( \frac{c_1}{c_2 c_3} \right) + \sigma_1 - \sigma_2 - \sigma_3
\end{cases}
\end{equation}
where $\sigma_1$, $\sigma_2$ and $\sigma_3$ are constant, related to the activity of ATP, ADP and P.

We consider the computational domain $\Omega = [0, 1]^2$ with periodic boundary conditions applied to all variables. The initial condition is given by
\begin{equation}
{\bm p}(x, y, 0) = \left( \cos(2\pi x), \sin(2\pi y) \right), \quad n(x, y, 0) = n_0 + n_1 \left( \sin(2\pi x) - \cos(2\pi y) \right),
\end{equation}
where $n_0 = 1$ and $n_1 = 0.3$. The initial concentrations of the chemical species are set as $c_1(x, y) = 1$, and $c_2(x, y) = c_3(x, y) = 0.01$. Model parameters are chosen as follows: $\nu_0 = 0.1$, $\lambda = 0.01$, $\Lambda = 10$, $\gamma = 1$, $\epsilon = 0.1$, $\sigma = (1, 0.1, 0.1)$, and $\eta = \eta_i = 0.1$ for all diffusion coefficients. We emphasize that these parameters are selected for illustrative and numerical purposes only; they are not based on physical measurements or calibration to experimental data.

We use the standard finite difference method with explicit Euler to solve the model. The grid size is $\Delta x = 1/50$, and the temporal stepsize is $\Delta t = 10^{-4}$.
We'll develop a structure-preseving numerical discretization for the variational model in feature work.

\begin{figure}[!htb]
\centering
\includegraphics[width = 0.75\textwidth]{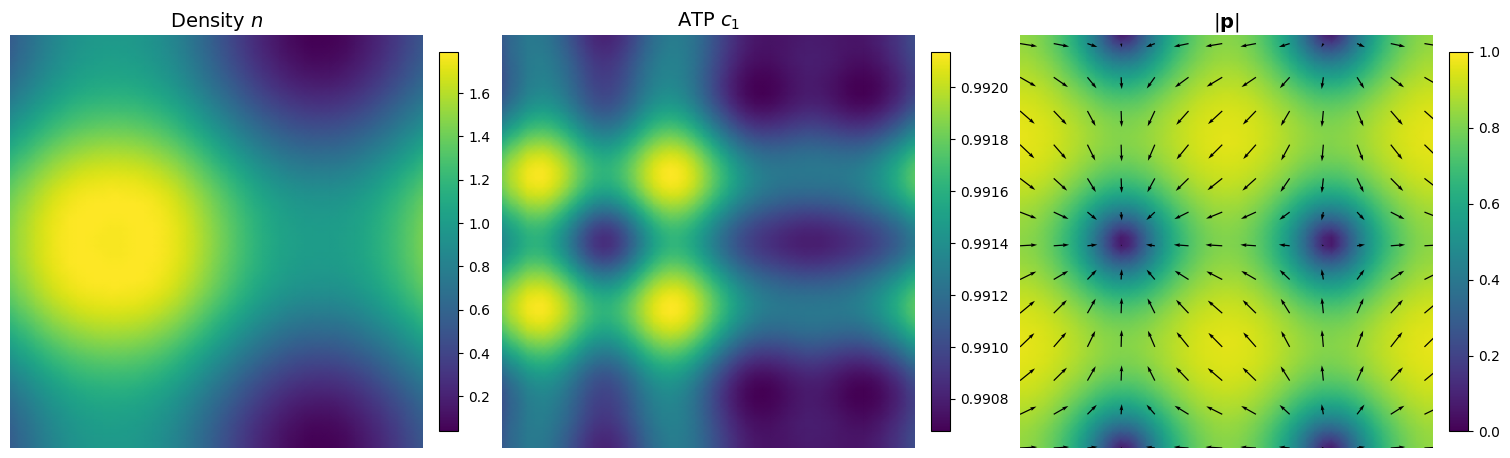}

\vspace{0.5em}
\includegraphics[width = 0.75\textwidth]{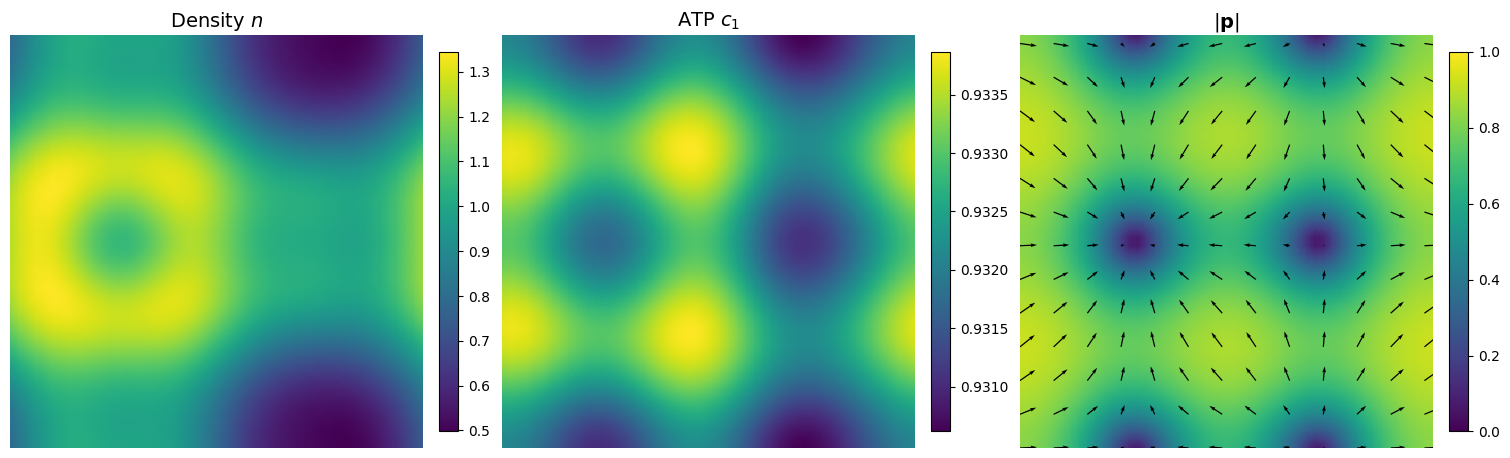}

\vspace{0.5em}
\includegraphics[width = 0.75\textwidth]{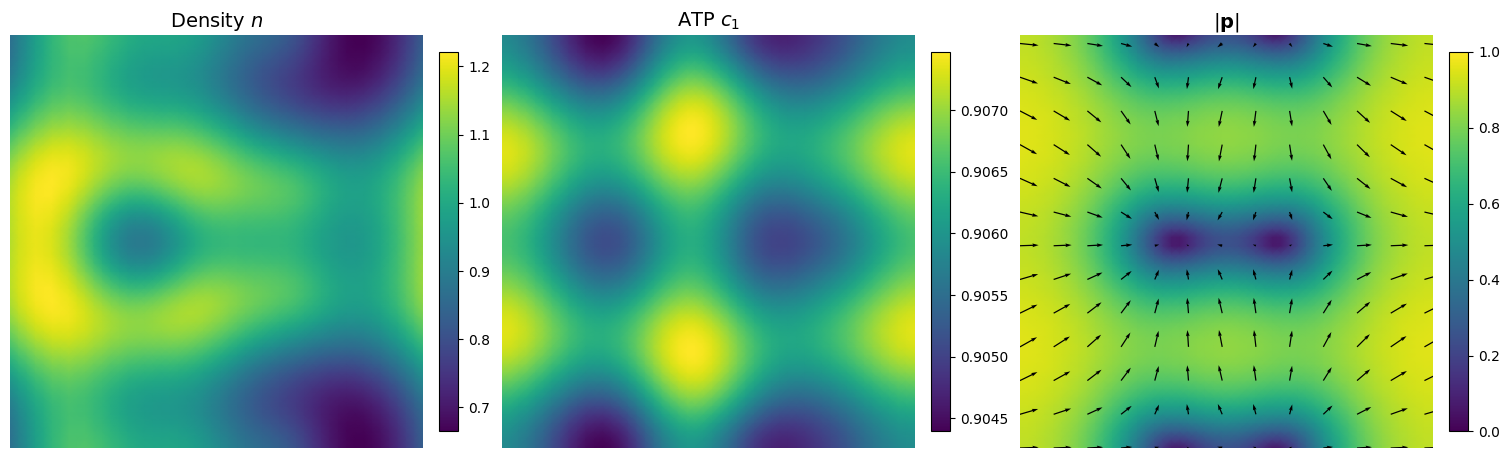}


\vspace{0.5em}
\includegraphics[width = 0.75\textwidth]{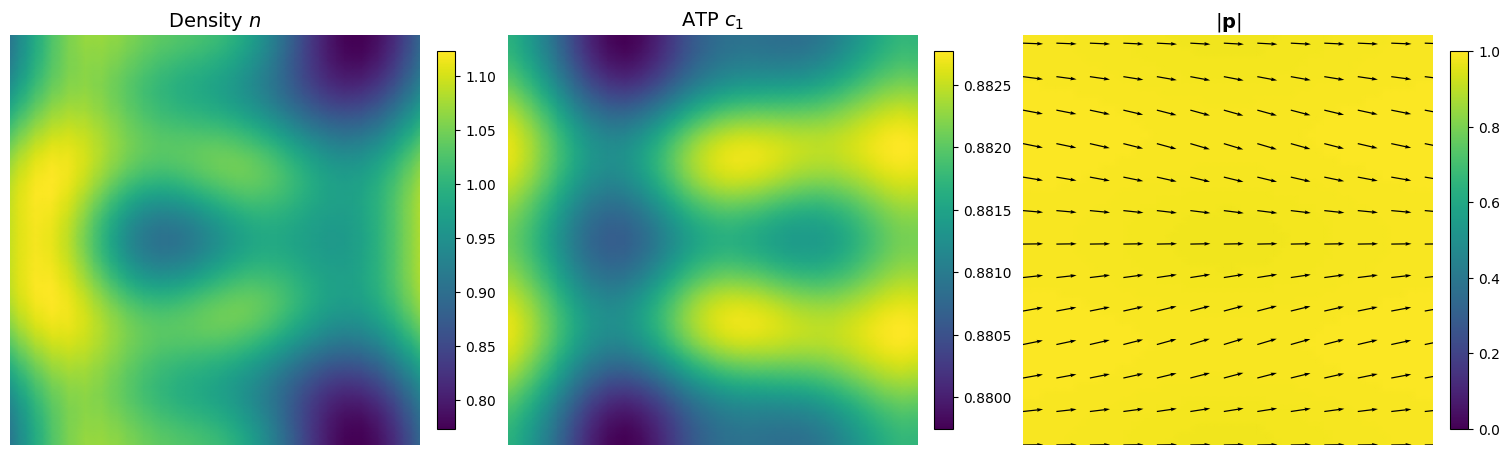}

\caption{Snapshots of the number density $n(\x, t)$, ATP concentration $c_1(\x, t)$, and polarization field ${\bm p}(\x, t)$ (colored by $|{\bm p}|$) at $t = 0.01$, $0.1$, $0.15$, and $0.2$ (from top to bottom).}\label{comp_result}
\end{figure}

Fig. \ref{comp_result} shows the simulation result, visualized by the number density $n(\x, t)$, the concentration of ATP $c_1({\bm x}, t)$, and the polarization field $\bm{p} (\x, t)$ (colored by $|{\bm p}|$).
Unlike passive nematic systems, the presence of activity enables the merging and annihilation of topological defects, allowing the system to escape from a quasi-equilibrium state and eventually reach a defect-free configuration. The free energy profile indicates that the system can remain in a metastable state for a significant duration, but sustained ATP hydrolysis provides the necessary energy input to drive the transition toward a lower-energy, defect-free state.

\begin{figure}[!h]
  \centering
\includegraphics[width= \textwidth]{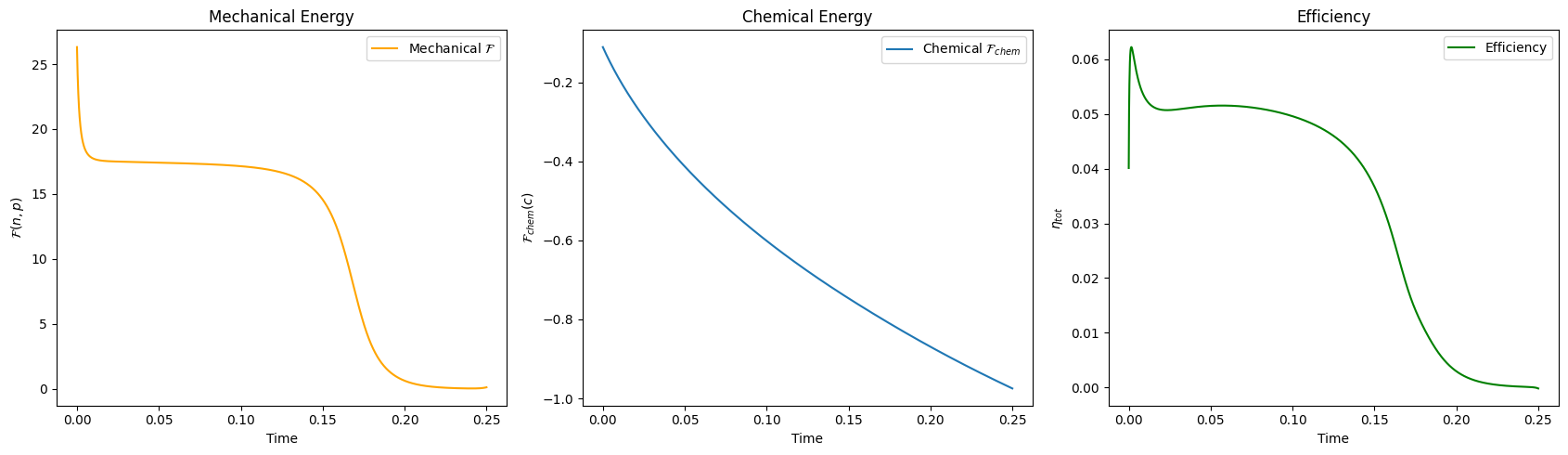}
\caption{Evolution of $\mathcal{F}(n, {\bm p})$, $\mathcal{F}_{\rm chem}({\bm c})$ and efficiency with respect to time.}\label{energy}
\end{figure}

Fig.~\ref{energy} shows the evolution of $\mathcal{F}(n, {\bm p})$, $\mathcal{F}_{\rm chem}({\bm c})$, and the efficiency $\eta_{\rm tot}$ over time. The results indicate that the system remains in a quasi-equilibrium state for a significant period before ATP-driven activity enables it to escape and reach a defect-free configuration with lower mechanical free energy. While $\mathcal{F}(n, {\bm p})$ generally decreases due to dissipation in both $n$ and ${\bm p}$, the efficiency plot demonstrates effective energy transduction from chemical to mechanical forms during the transition. Moreover, it can be noticed that $\eta_{\rm tot}$ remains relatively high throughout the defect-annihilation process, highlighting the sustained conversion of chemical energy into mechanical work.

\section{Conclusions}

In this work, we developed a thermodynamically consistent extension of the simplified Toner--Tu model by explicitly incorporating ATP hydrolysis as the energetic source of activity. 

Our model is constructed using an energetic variational approach within a nonequilibrium thermodynamic framework, ensuring that the total free energy, comprising both mechanical and chemical components, monotonically decreases over time, consistent with the first and second laws of thermodynamics. The formulation captures a two-way coupling between chemical and mechanical processes: the reaction rate of ATP hydrolysis is influenced by mechanical alignment, while active transport is, in turn, modulated by the local reaction rate. Although the full system is dissipative, the free energy of the mechanical subsystem may increase, and the mechanical dynamics can exhibit out-of-equilibrium behavior due to the transduction of chemical energy into mechanical work. In the absence of ATP consumption, the model naturally reduces to a passive system, providing a clear energetic interpretation of activity.

We also analyzed the transduction of chemical energy into mechanical work, identifying the conditions under which ATP hydrolysis fuels transport and alignment. A simple expression for energy conversion efficiency was derived and used to characterize the system's behavior under different parameter regimes.  We also present numerical simulations demonstrating how ATP consumption drives the merging of topological defects and enables the system to escape a quasi-equilibrium, which can not be observed in passive nematics. In future work, we plan to conduct a more detailed mathematical analysis of solution behavior, perform comprehensive numerical investigations across broader parameter regimes, and develop structure-preserving discretization schemes that better reflect the underlying variational principles.

This work provides a variational and physically grounded framework for studying active matter powered by biochemical reactions. The chemo-mechanical coupling we introduced can be extended to more complex settings, including hydrodynamic interactions, anisotropic geometries, and micro-macro coupling. It also opens new doors for investigating energy transduction and efficiency in biologically inspired active matter systems.

\section*{ACKNOWLEDGMENTS}
This work is partially support by NSF DMS-2410740. The author would like to thank Professor Chun Liu for his insightful guidance and many stimulating discussions.


\bibliographystyle{siam}
\bibliography{Ref_Active}

\end{document}